\documentstyle[12pt,wrapfig]{article}
\author{Leonid Didukh, Oleksandr Kramar, Yuriy Skorenkyy, Yuriy
Dovhopyaty \\ Ternopil State Technical University, Department of
Physics}

\date{}
\title{Magnetic field dependence of conductivity and effective mass of
carriers in a model of Mott-Hubbard material}

\begin{document}
\maketitle

\begin{abstract}
The influence of external magnetic field $h$ on a static
conductivity of Mott-Hubbard material which is described by model
with correlated hopping of electrons has been investigated. By
means of canonical transformation the effective Hamiltonian which
takes into account strong intra-site Coulomb repulsion and
correlated hopping is obtained. Using a variant of generalized
Hartree-Fock approximation the single-electron Green function and
quasiparticle energy spectrum of the model have been calculated.
The static conductivity $\sigma$ has been calculated as a function
of $h$, electron concentration $n$ and temperature $T$. The
correlated hopping is shown to cause the electron-hole asymmetry
of transport properties of narrow band materials.

keywords{Mott-Hubbard material, conductivity, magnetic field}

pacs{72.15-v; 72.80.Ga.}
\end{abstract}

The achievements of the recent years in the field of strongly
correlated electron  systems give us the opportunity to understand
the properties of narrow-band materials, in particular those in
which metal-insulator transition under the action of external
influences (pressure, doping, temperature) is realized~\cite{imad98}.
The strongly correlated electron systems
demonstrate unusual transport properties~\cite{mott90}. For
understanding of the physical mechanisms, which cause these
peculiarities, the experimental and theoretical researches of
the temperature dependence of conductivity are needed. The results
concerning low-frequency behavior of conductivity are of the prior
importance, because it gives the information about the scattering
processes close to the Fermi surface. The theoretical
investigation of conductivity $\sigma (\omega,T)$ are mainly
concentrated in the limit $T=0$. Behavior of the static
conductivity $\sigma (T)=\sigma(\omega=0,T)$ at $T>0$ has not been
studied sufficiently.

Theoretical investigations of the optical conductivity in the
Hubbard model~\cite{hubb1} in the frameworks of the Kubo linear
response theory~\cite{kubo57} last for many decades, we note here
the investigations by analytical methods: moment
method~\cite{ohata70}, in composite operators method~\cite{manc},
in the mean-field theory~\cite{hirs99,hirs00}, in the perturbative
theory method~\cite{bren92,wermb93} in the limit of weak
interaction ($|t| \gg U$), in method of the memory
function~\cite{goet72,plak96}, in the opposite limit ($U \gg |t|$).
The conductivity has been intensively studied in one-dimensional
Hubbard model~\cite{fye90}-\cite{Steph90}, where the numerical
results can be compared with exact ones obtained by Bethe ansatz
application.

For the numerical investigation of the conductivity in Hubbard model
the exact diagonalization of finite clusters has been used~\cite{dag94}
for 4 x 4 sites cluster,  quantum Monte-Carlo method for 8, 10
sites~\cite{moreo90}, 3 x 3 sites~\cite{Steph90},
 8 x 8 sites~\cite{scal92},  12 x 12 sites~\cite{bulu94}.
The investigations by numerical methods were carried out mainly for
two-dimensional lattice (this is caused by the interest to the
high-temperature superconductivity phenomena~\cite{plak} in systems
with $CuO$ planes), the conductivity of three-dimensional system has been
studied only in narrow interval close to half-filling at weak and
intermediate interactions~\cite{tan92}.

The optical conductivity has been studied also by the dynamical
mean field theory (DMFT) in the limit of infinite spatial
dimension~\cite{geor96}. Different DMFT equations solvers were
used: iterated perturbation theory IPT~\cite{geor93}, non-crossing
approximation NCA~\cite{prus93}, second order perturbation theory
2OPT~\cite{roz95}. The authors have considered the symmetrical Hubbard
model at half-filling close to metal-insulator transition
(intra-atomic Coulomb repulsion $U \simeq 2w$, where $2w$ is the
band width) and have obtained a good agreement with experimental
data for some Mott-Hubbard systems. But for realistic models the
non-local (dependent on wave vector) contributions to self-energy
and transport characteristics are important. Besides, investigations
in the framework of DMFT consider mainly the case of half-band filling.

In works~\cite{2}-\cite{did97jps} the essential importance of
taking into account in the Hubbard model the correlated hopping of
electrons and narrow-band model with non-equivalent
Hubbard subbands has been proposed. In such a model the hopping
integrals which describe translation movement of holes and
doublons, differ one from the other and from the activation
processes integral. Similar models have been studied intensively in
recent years~\cite{5}-\cite{7}. For the generalized model which
take into account the non-equivalency of Hubbard subbands,
reliable results for conductivity do not exist, so the analytical study
of conductivity in the framework of the realistic models of the
electronic subsystem is of importance.

In this work we
show that application of the variant of projection procedure for
Green function allows to reproduce some peculiarities of static
conductivity of narrow-band material in the limit of strong Coulomb
correlation and investigate the effect of external influences
like temperature change, doping, pressure and magnetic field. We
apply our approach to the Hamiltonian, which, besides of intra-site
Coulomb repulsion $U$, strong in comparison with inter-site
hopping $t_{ij}$, additionally describes correlated hopping of
electrons (influence of electron concentration  $n$ on the hopping
processes) and show that it leads to the electron-hole asymmetry
of conductivity and other characteristics.

We write the Hamiltonian of correlated electron system in
representation of $X_i^{kl}$ Hubbard operators:
\begin{eqnarray}\label{ham1}
H=H_0+H_1+H'_1+H_{ex},
\end{eqnarray}
where
\begin{eqnarray*}
&&H_0=-\mu \sum_{is} \left(X_i^{s}+X_i^2\right)+
U\sum_{i}X_i^2
+{1\over 2}NV_0\kappa u^2 +\mu_B h \sum_{is} \eta_{s}X_i^{s},\\
&&H_1={\sum \limits_{ijs}}'t_{ij}(n)X_i^{s 0} X_j^{0s} + {\sum
\limits_{ijs}}'\tilde{t}_{ij}(n)X_i^{2s}X_j^{s 2},
\\
&&H'_1={\sum \limits_{ijs}}' \left(t'_{ij}(n) \eta_{s}X_i^{s
0}X_j^{\bar{s} 2}+h.c.\right).
\\
&&H_{ex}=-{1\over 2}{\sum\limits_{ijs}}'J(ij)\left(
\left(X_i^{s}\!+\!X_i^{2}\right)\left(X_j^{s}\!+\!X_j^{2}\right)+
X_i^{s\bar{s}}X_j^{\bar{s}s}\right).
\end{eqnarray*}
Here operator $X_i^{kl}$ describes transition of site $i$ from
state $|l\rangle$ to state $|k\rangle$, $\mu$ is the chemical
potential, $U$ denotes the energy of
intra-site Coulomb repulsion of electrons, $J$ stands for the
direct inter-site exchange interaction, $\kappa$ is the elastic
constant, $V_0$ is the initial volume of the crystal, $N$ is the
number of lattice sites, $\mu_B$ is Bohr magneton, $h$ stands for
the external magnetic field, $\eta_s=1$ if $s=\uparrow$ and $-1$
otherwise. Translation processes are characterized by different
hopping integrals, namely $t_{ij}(n)=(1+\alpha u)(1-\tau_1
n)t_{ij}$ and $\tilde{t}_{ij}(n)=(1+\alpha u)(1-\tau_1
n-2\tau_2)t_{ij}$ are hopping parameters for holes and doublons,
respectively; $t'_{ij}(n)=(1+\alpha u)(1-\tau_1 n-\tau_2)t_{ij}$
is parameter of the hopping of an electron between doublon and
hole; correlated hopping parameters $\tau_2$ and $\tau_1$ describe
the influence of sites involved into the hopping process and
neighbor sites, respectively; parameter $\alpha<0$ takes into
account the renormalization of bandwidth $2w=2z|t_{ij}|$ at strain
$u$~\cite{stas}, z is the number of nearest neighbor to a site.

We restrict ourselves to considering the strong correlation limit,
namely $U\gg w(n)$. In such a system at partial filling of the band
the conductance is mainly due to electron hopping processes within
the Hubbard subbands and interband hopping can be neglected. At
these condition we apply the canonical
transformation~\cite{did97jps} to the Hamiltonian~(\ref{ham1}).
\begin{eqnarray}
H_{eff}=e^SHe^{-S},
\end{eqnarray}
where
\begin{eqnarray}
S=\sum_{ij}\left(L(ij)\left(X_i^{\uparrow 0}X_j^{\downarrow 2}
-X_j^{\downarrow 0}X_i^{\uparrow 2}\right)-h.c.\right),
\end{eqnarray}
with $L(ij)=\frac{t'_{ij}(n)}{U}$.
The operator $S$ is taken to exclude the processes with pair hopping
of holes and doublons in the first order in hopping parameter:
\begin{eqnarray}
H'_1+[S;H_0]=0.
\end{eqnarray}

Finally, we obtain the effective Hamiltonian:
\begin{eqnarray}
H_{eff}=H_0+H_1+H_{ex}+\tilde{H}_{ex},
\end{eqnarray}
where
\begin{eqnarray}
\tilde{H}_{ex}=-\frac{1}{2}{\sum\limits_{ijs}}'
\tilde{J}(ij)\left(X_i^{s}X_j^{\bar{s}}-
X_i^{s\bar{s}}X_j^{\bar{s}s}\right)
\end{eqnarray}
with the indirect exchange interaction parameter
$\tilde{J}(ij)=\frac{(t'_{ij}(n))^2}{U}$.

 Using the projection procedure~\cite{apr} for the case of $n<1$ we obtain
for the single particle energy spectrum:
\begin{eqnarray}
\label{spectr1} E_s({\bf k})={-\mu}-zJ n_\sigma-z\tilde{J} n_{\bar{\sigma}}
+\alpha_st_{\bf k}(n)+\beta_s,
\end{eqnarray}
where the correlated narrowing of the band and spin-dependent shift of
subband center are
\begin{eqnarray}
&&\alpha_s={2-n +\eta_s m\over 2}+{n^2-m^2 \over
2(2-n+\eta_s m)},\\
&&\beta_s=-{2\over (2-n+\eta_s m)}{\sum \limits_{\bf k}t_{\bf
k}(n)\langle X^{\bar{s}0}_{i}X^{0\bar{s}}_{j}\rangle_{\bf k}},
\end{eqnarray}
respectively, and $z$ is the number of nearest neighbors to a site.
The respective results for $n>1$:
\begin{eqnarray}
\label{spectr2} \tilde{E}_s({\bf k})={-\mu}+U
-zJ n_\sigma-z\tilde{J} n_{\bar{\sigma}}
+\tilde{\alpha}_s\tilde{t}_{\bf k}(n)+\tilde{\beta}_s,
\end{eqnarray}
\begin{eqnarray}
&&\tilde{\alpha}_s={n +\eta_s m\over 2}+{n^2-m^2 \over
2(n+\eta_s m)},\\
&&\tilde{\beta}_s=-{2\over (n+\eta_s m)}{\sum \limits_{\bf k}
\tilde{t}_{\bf k}(n)\langle X^{s2}_{i}X^{2s}_{j}\rangle_{\bf k}},
\end{eqnarray}
describe upper Hubbard subband of halfbandwidth
$\tilde{w}(n)=z|\tilde{t}_{ij}(n)|$.

With use of the method of works~\cite{bari,pre}, we calculate the
$xx$-component of static electronic conductivity
$\sigma_{xx}=\sigma+\tilde{\sigma}$, where
\begin{eqnarray}
\label{sig1} \sigma=-\frac{e^2\tau z}{2Na}\sum_{ijs}\langle X_i^{s
0} X_j^{0s}\rangle t_{ij}(n),
\end{eqnarray}
is the conductivity of lower $(0-s)$-subband,
\begin{eqnarray}
\label{sig2} \tilde{\sigma}=-\frac{e^2\tau
z}{2Na}\sum_{ijs}\langle X_i^{2s } X_j^{s2}\rangle
\tilde{t}_{ij}(n) ,
\end{eqnarray}
is the conductivity of upper $(\uparrow\downarrow-\bar{s})$-subband.

\begin{figure}[htbp]
\begin{minipage}[t]{75mm}
\begin{picture}(65,65)
\unitlength 1mm
\put(-5,75){\special{em:graph 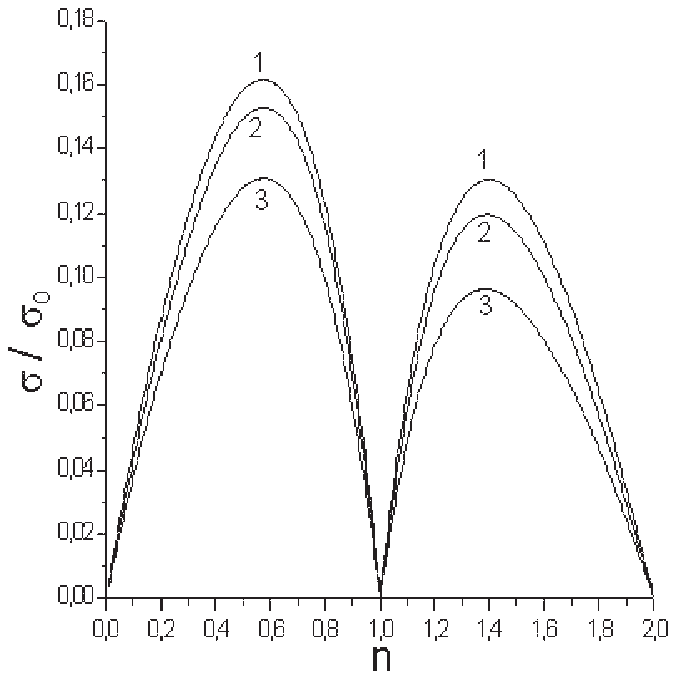}}
\end{picture}
\caption{The concentration dependencies of the conductivity
at correlated hopping $\tau_1=\tau_2=0.1$ in absence of the
external magnetic field and exchange interactions:
curve 1 corresponds to temperature $\Theta/w=0.01$,
curve 2 -- to $\Theta/w=0.1$, curve 3 -- to $\Theta/w=0.2$.}
\label{sig_n}
\end{minipage}
\hfill
\begin{minipage}[t]{75mm}
\unitlength 1mm
\begin{picture}(65,65)
\put(0,75){\special{em:graph sig_nh.eps}}
\end{picture}
\caption{The concentration dependencies of the conductivity in the
external magnetic field at $\Theta/w=0.02$, $zJ_{eff}/w=0.02$,
$\tau_1=\tau_2=0$:
curve 1 corresponds to $h/w=0$,
curve 2 -- to $h/w=0.005$, curve 3 -- to $h/w=0.02$.}
\label{sig_nh}
\end{minipage}
\end{figure}

\begin{figure}[htbp]
\begin{minipage}[t]{75mm}
\begin{picture}(65,65)
\unitlength 1mm
\put(-5,75){\special{em:graph 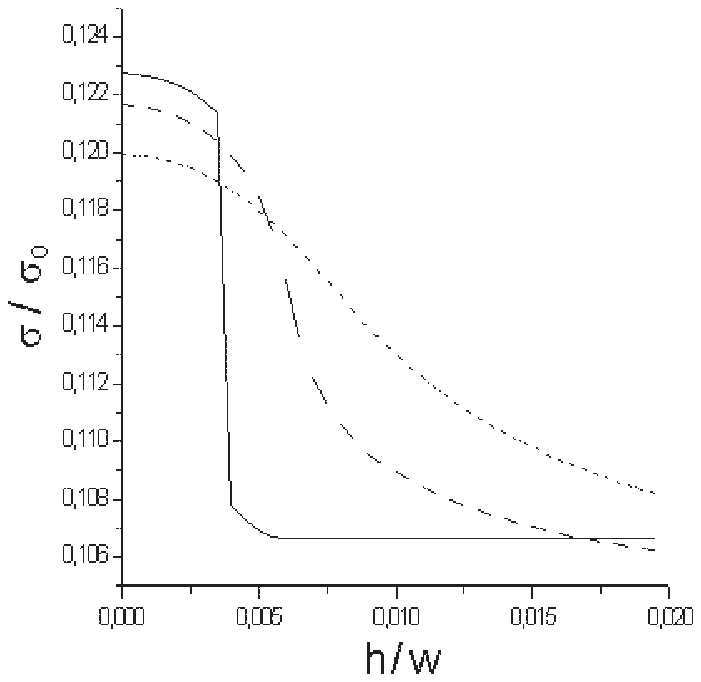}}
\end{picture}
\caption{The dependencies of the conductivity on magnetic field
for the electron concentration $n=0.3$ and $zJ_{eff}/w=0.01$,
$\tau_1=\tau_2=0$:
solid curve corresponds to $\Theta/w=0.02$,
dashed curve -- to $\Theta/w=0.04$,
dashed-dotted curve  -- to $\Theta/w=0.06$.}
\label{sig_h}
\end{minipage}
\hfill
\begin{minipage}[t]{75mm}
\unitlength 1mm
\begin{picture}(65,65)
\put(0,75){\special{em:graph 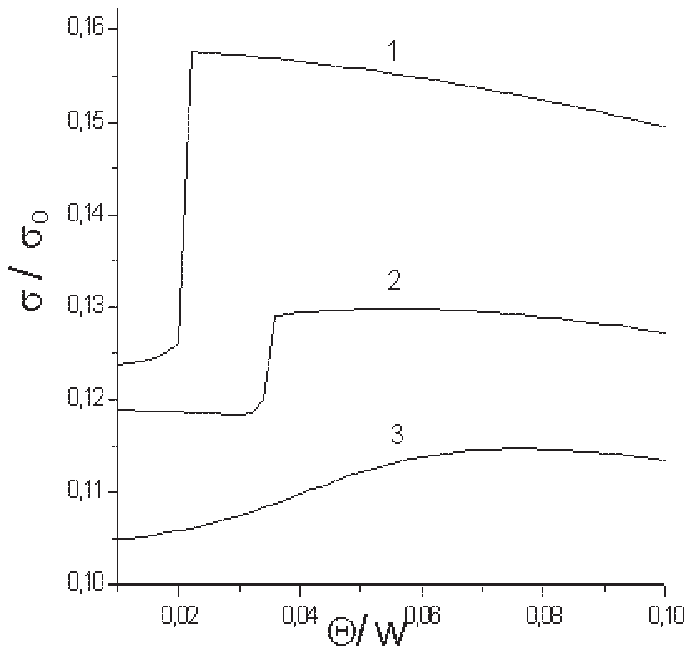}}
\end{picture}
\caption{The dependencies of the conductivity on temperature
in the external magnetic field $h/w=0.01$ at $zJ_{eff}/w=0$ and
$\tau_1=\tau_2=0$:
curve 1 corresponds to $n=0.3$,
curve 2 -- to $n=0.35$, curve 3 -- to $n=0.45$.}
\label{sig_tet}
\end{minipage}
\end{figure}

The magnetization can be calculated from the equation
\begin{eqnarray}
\label{rivn_m_T}
\exp\left(\frac{2h+\beta_\downarrow-
\beta_\uparrow+zJ_{eff}m}{\Theta}\right)
=
\frac{\sinh\left( \frac{w(n)\alpha_\downarrow}{\Theta}
\frac{1-n}{1-n_\uparrow}\right)
}
{\sinh\left( \frac{w(n)
\alpha_\uparrow}{\Theta}\frac{1-n}{1-n_\downarrow}\right)}
\frac{
\sinh \left( \frac{w(n)\alpha_\uparrow}{\Theta}
\frac{n_\uparrow}{1-n_\downarrow}
\right)
}
{
\sinh \left( \frac{w(n)\alpha_\downarrow}{\Theta}
\frac{n_\downarrow}{1-n_\uparrow}
\right)
}
\end{eqnarray}
for $n<1$ and corresponding equation in which the substitutions
$n \to 2-n$, $w(n) \to \tilde w(n)$, $\alpha_s \to \tilde \alpha_s$,
$\beta_s \to \tilde \beta_s$ are made. Here $J_{eff}=J-\tilde{J}$.
Using these expressions we have numerically calculated the static
conductivity $\sigma_{xx}$ as a function of electron concentration
(Figs.~\ref{sig_n},~\ref{sig_nh}), temperature (Fig.~\ref{sig_tet}),
and magnetic field (Fig.~\ref{sig_h}).

\begin{figure}[htbp]
\begin{minipage}[t]{75mm}
\begin{picture}(65,65)
\unitlength 1mm
\put(-5,75){\special{em:graph 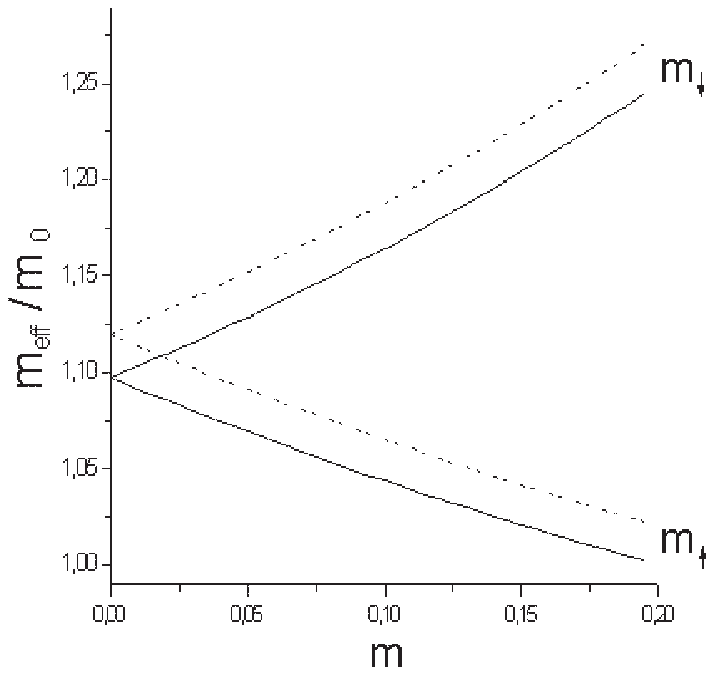}}
\end{picture}
\caption{The dependence of current carriers effective mass
$m_{eff}/m_0$ on the magnetization at $n=0.2$, $\Theta/w=0.02$,
$zJ_{eff}/w=0$:
solid curves correspond to $\tau_1=\tau_2=0$;
dashed curves -- to $\tau_1=\tau_2=0.1$.}
\label{m_m}
\end{minipage}
\hfill
\begin{minipage}[t]{75mm}
\unitlength 1mm
\begin{picture}(65,65)
\put(-5,75){\special{em:graph 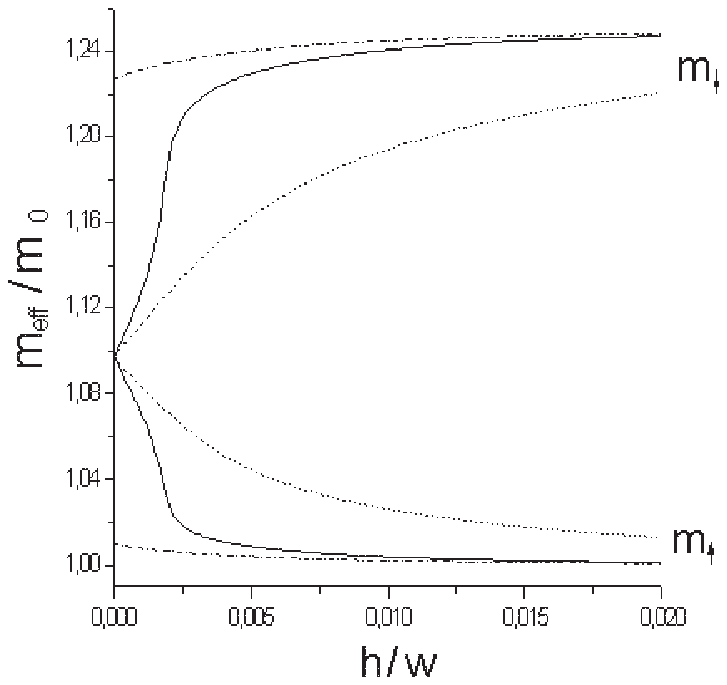}}
\end{picture}
\caption{The dependence of current carriers effective mass
$m_{eff}/m_0$ on the applied magnetic field:
solid curves correspond to $zJ_{eff}/w=0$, $\Theta /w=0.02$;
dotted curves -- to $zJ_{eff}/w=0$, $\Theta/w=0.05$;
dash-dotted curves -- to $zJ_{eff}/w=0.05$, $\Theta/w=0.02$.}
\label{m_h}
\end{minipage}
\end{figure}

From Fig.~\ref{sig_n} one can see that in the considered model
with correlated hopping of electrons the conductivity provided by
carriers in upper subband is lower than the conductivity, provided
by carriers from upper subband. This effect was discussed in
work~\cite{pre}, it is a manifestation of the electron-hole
asymmetry, inherent to real transition metal compounds. Other
important feature is the change of current carrier type from
metallic to semiconducting type in the vicinity of $n=2/3,4/3$ and
from semiconducting type to metallic one at $n=1$. Increasing the
correlated hopping we shift maxima of the conductivity closer to
the half-filling. External magnetic field changes the
concentration dependence of $\sigma$ qualitatively (see
Fig.~\ref{sig_nh}). The higher is the electron concentration, the
less pronounced is the effect of applied magnetic field. At small
concentration of electrons the band is fully polarized, in such a
ferromagnetic system the conductivity is considerably lower than
in paramagnetic state. If electron concentration increase, the
decrease of magnetization leads to the increase of conductivity,
$\sigma$ approaches its value in paramagnetic state. At the same
time, due to the changes in the correlation band narrowing and the
shift of subband center, the position of conductivity maximum
changes from $n=2/3$ in saturated ferromagnetic state
(Fig~\ref{sig_nh}, 2) to $n=0.5$ in ferromagnetic state
(Fig~\ref{sig_nh}, 3). The change of conductivity with increase
of the applied field can be very sharp at low temperature,
(Fig~\ref{sig_h}, solid curve). The increase of temperature leads
to the decrease of $\sigma$ and its changes become more smooth.
Plateau of $\sigma(h)$ dependence signifies, that at low
temperatures the fully polarized state is reached in very low
field. Much higher field is needed to polarize spins at higher
temperatures (Fig~\ref{m_h}, long-dashed and short dashed
curves).
Sharp increases of $\sigma$ with increase of temperature is related to
the transition from the polarized ferromagnetic state to paramagnetic
one, in each state the conductivity decreases with rise of temperature.
Very small changes of electron concentration can greatly affect the
temperature dependence of conductivity (see Fig.~\ref{sig_tet}).

In the framework of the considered model one can naturally introduce
the notions of "wide" (lower) and "narrow" (upper) energy bands and,
correspondingly, "light" and "heavy" current carriers with the effective
masses
\begin{eqnarray}
m^*_s=\left({\partial ^2E_s(\bf k)\over \partial {\bf k}^2}\right)^{-1},
\qquad
\tilde{m}^*_s=\left({\partial ^2\tilde{E}_s(\bf k)\over \partial {\bf k}^2}
\right)^{-1},
\label{m*}
\end{eqnarray}
where $E_s(\bf k)$ is the energy spectrum of current carriers in lower
($s-0$) band, $\tilde{E}_s(\bf k)$ is the energy spectrum of current
carriers in upper ($\bar{s}-\uparrow\downarrow$) subband.

We have obtained that effective mass of heavy carriers can increase
substantially with the increase of electron concentration.
Because of the correlated hopping the heavy carriers can be realized
in lower ($s-0$) subband as well.
It is worthwhile to note that the notion "effective mass of
current carrier" has conditional sense, different from the
standard one, used in band theory. The definitions~(\ref{m*}) are
related to the expressions for band spectrum which describe the
transitions between $|s\rangle$- and $|0\rangle$-states as well as
$|\!\uparrow\downarrow\rangle$- and $|\bar{s}\rangle$-states. Then
$m^*$ and $\tilde{m}^*$ are effective masses of respective
transitions; for the cases when subbands are almost empty or
almost full, $m^*$ and $\tilde{m}^*$ can be interpreted as
effective masses of electronic and hole states.
\begin{wrapfigure}[20]{r}{80mm}
\begin{minipage}[b]{75mm}
\unitlength 1mm
\begin{picture}(70,70)
\put(-7,70){\special{em:graph 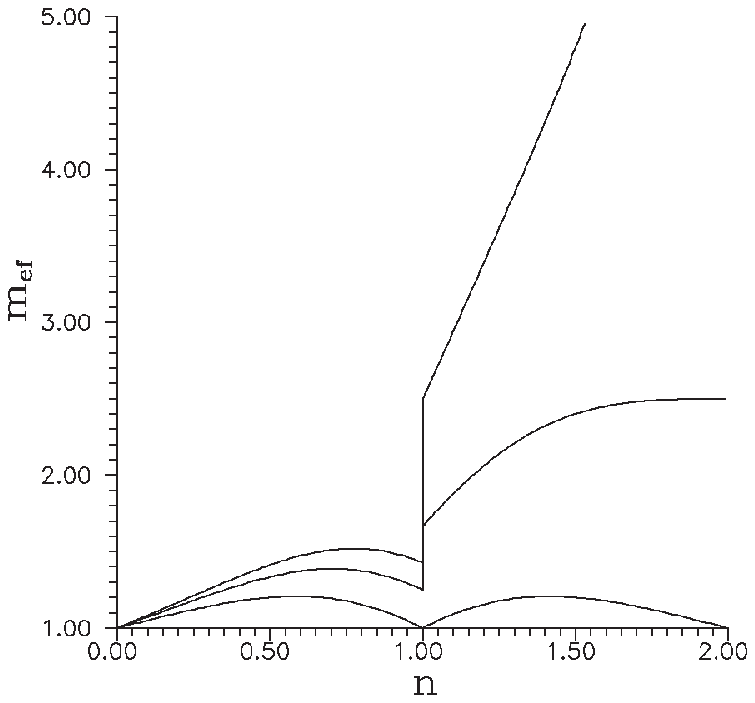}}
\end{picture}
\caption{The dependence of current carriers effective mass
$m_{eff}/m_0$ on electron concentration:
upper curves correspond to $\tau_1=\tau_2=0.3$;
middle curves -- to $\tau_1=\tau_2=0.2$;
lower curves -- to $\tau_1=\tau_2=0$.}
\label{m_eff}
\end{minipage}
\end{wrapfigure}
In the paramagnetic state for the case $n\ll 1$ the value
$m^*_1={\hbar\over 2a^2 |t(n)|}\simeq{\hbar\over 2a^2 |t|}$
can be identified as an effective mass of $|s\rangle$-states,
i.e. electrons, for the case  $n=1-\varepsilon$ ($\varepsilon\ll 1$)
the value
$m^*_2={\hbar\over 2a^2 |t(n)|}|_{n=1}$
is an effective mass of a hole.
If $n>1$, then for $n=1+\varepsilon$ we have translational motion of
doublons ("extra electrons") with masses
$\tilde{m}^*_1={\hbar\over 2a^2 |\tilde{t}(n)|}|_{n=1}$,
and for $n=2-\varepsilon$ we have the effective mass of holes
$\tilde{m}^*_2={\hbar\over 2a^2 |\tilde{t}(n)|}|_{n=2}$.
From these results one can see that the effective mass
of carrier in upper band can differ substantially from the effective
masses in the case when the conductivity is due to $s-0$-transitions.
It is important to note that passing from the regime of conductivity
provided by the carriers in lower band to the regime when it is provided
by $\bar{s}-\uparrow\downarrow$-transitions, the effective mass increase
stepwise at the point $n=1$ (the behavior of the effective
mass dependence on the electron concentration is shown in Fig.~\ref{m_eff}).

The effective mass dependence on the magnetization and external
magnetic field is shown in Figs.~\ref{m_m} and \ref{m_h},
respectively. Rise of magnetization leads to the rise of
difference in effective masses of spin-up and spin-down current
carriers (Fig.~\ref{m_m}). The slope of the dependencies
$m_{eff}(m)$ changes with the rise of $m$, too. This leads to the
decrease of overall transport in ferromagnetic state, though
effective masses of carrier with majority becomes lower. The
results shown in Fig.~\ref{m_m} qualitatively agree with the
corresponding plot of work~\cite{spal90}, where Gutzwiller
approximation has been used to calculate effective masses of
current carriers. The correlated hopping, favoring localization,
shift effective masses up. Different possible scenarios of
$m_{eff}(h)$ dependence are shown in Fig.~\ref{m_h}. At high
temperature $m_{eff}$ changes monotonically (Fig.~\ref{m_h},
dotted curve) while at low temperature system in unstable towards
the transition to polarized state (Fig.~\ref{m_h}, solid curve).
Direct exchange interaction can stabilize ferromagnetically
polarized state in the less then half-filled band even in weak
magnetic field (Fig.~\ref{m_h}, dash-dotted curve).
In this paper we have used a model with correlated hopping of
electrons to study the influence of external magnetic field,
temperature and doping on a static conductivity of Mott-Hubbard
material. In the regime of strong Coulomb interaction
hybridization of the Hubbard subbands do not contribute
essentially to the transport properties so we have excluded it
from the effective Hamiltonian by means of canonical
transformation. The single-electron Green function and
quasiparticle energy spectrum of the model have been calculated
using a variant of generalized Hartree-Fock approximation. This
procedure has allowed us to obtain analytical expression for the
band narrowing factor and relatively simple and transparent
equations for calculation of spin-dependent shift of the band
center, magnetization, static conductivity and effective mass of
the current carrier. The results of our study generalize the
results of works~\cite{manc,dag94}, where the static conductivity
of the Hubbard model has been calculated as a function of electron
concentration $n$, for a wider class of systems, for which the
correlated hopping should be taken into account. In the limiting
case of the absence of correlated hopping our results agree with
the concentrational dependencies obtained in the composite
operator method~\cite{manc}, exact diagonalization~\cite{dag94}
and Monte-Carlo simulations~\cite{scal92} results. The temperature
dependence of conductivity in paramagnetic state, calculated in
this work, agrees with the corresponding results of
DMFT~\cite{prus93,uhri95}. In the magnetic field static
conductivity reflects the changes of single electron energy
spectrum through correlation narrowing of the band and shift of
subband center. The temperature and concentration dependencies  of
$\sigma$ are governed by changes of system magnetization in
external magnetic field. We have found that in the ground state
the saturated ferromagnetic state is stable while at non-zero
temperature magnetization has the concentration dependence which
agrees with work~\cite{ok05}. Such behavior of magnetization leads
to $\sigma(n)$ dependence with maxima at quarter and three-quarter
fillings in distinction from paramagnetic ones, obtained in
work~\cite{pre}. At non-zero temperatures the sharp changes of
$\sigma(n)$ dependence are possible. It is due to complicated
character of temperature dependence of band narrowing factor.
Effective mass of quasi-particles appear to be spin-dependent and
substantially varies with magnetic field. These results are in
agreement with the analysis of work~\cite{spal90} and experimental
data of work~\cite{flou05} for heavy-fermion compounds. Taking
into account the correlated hopping which is inherent to real
narrow band materials allows us to describe electron-hole
asymmetry of these processes which is observed in real materials.

\section*{Acknowledgement}
Authors are grateful to Prof. I.V. Stasyuk, Prof. J. Spa\l ek and
Dr. A.M. Shvajka for the enlightening discussions. This work was
supported by Ukrainian Fund for Fundamental Research under grant
No 02.07/266.


\begin{thebibliography}{99}
\bibitem{imad98} Imada M., Fujimori A., and Tokura Y.,
 Rev. Mod. Phys. {\bf 70}, 1039 (1998).
\bibitem{mott90}Mott N.~F. Metal-insulator transition.- London: Taylor \&
Francis, 1990.- 286 p.
\bibitem{hubb1} Hubbard J., Proc.Roy. Soc. {\bf A276} 238 (1963).
\bibitem{kubo57} Kubo~R. J., Phys. Soc. Jpn. {\bf 12} 570 (1957).
\bibitem{ohata70} Ohata~N. and Kubo~R., J. Phys. Soc. Jpn. {\bf 28} 1402
(1970).
\bibitem{manc} Mancini~W. and Villiani~D., Phys. Let.~A {\bf 261}
357 (1999).
\bibitem{hirs99} Hirsch J.~E., Phys. Rev. B. {\bf 59} 6256 (1999).
\bibitem{hirs00} Hirsch J.~E., Phys. Rev. B. {\bf 62} 14131 (2000).
\bibitem{bren92} Brenig~W., Z. Phys. {\bf 89} 187 (1992).
\bibitem{wermb93} Wermbter~S. and Tewordt~L., Physica C . {\bf 211}
132 (1993).
\bibitem{goet72} Goetze W., Woelfle P., Phys. Rev. B.
{\bf 6} 1226 (1972).
\bibitem{plak96} Plakida N., J. Phys. Soc. Jpn.
{\bf 65} 3964 (1996).
\bibitem{fye90} Fye~R., Martins~M. and Scalettar~R.~T.,
Phys. Rev. B. {\bf 42} 6809 (1990).
\bibitem{fye91} Fye~R. et al., Phys. Rev. B.
{\bf 44} 6909 (1991).
\bibitem{Steph90} Stephan~W. and Horsh~P., Phys. Rev. B. {\bf 42} 8736 (1990).
\bibitem{dag94} Dagotto E., Rev. of Mod. Phys. {\bf 66} 763 (1994).
\bibitem{moreo90} Moreo~A. and Dagotto~E.,
Phys. Rev. B. {\bf 42} 4786 (1990).
\bibitem{scal92}  Scalapino D. J., White S. R., Zhang  S. C.,
Phys. Rev. Lett. {\bf 68} 2830 (1992).
\bibitem{bulu94} Bulut N., Scalapino D.~J., White S.~R.,
Phys. Rev. Lett. {\bf 73} 748 (1994).
\bibitem{plak} Plakida~N.~M. High-Temperature Superconductivity.-
Berlin: Springer-Verlag, 1995.
\bibitem{tan92}  Tan~L. and Callaway~J., Phys. Rev. B.
{\bf 46} 5499 (1992).
\bibitem{geor96} Georges A., Kotliar G., Krauth W., Rozenberg M.,
Rev. Mod. Phys. {\bf 68} 13 (1996).
\bibitem{geor93} Georges A., Krauth W.,
Phys. Rev.~B. {\bf 48} 7167 (1993).
\bibitem{prus93} Pruschke~Th., Cox~D.~L., Jarrel~M.,
Europhys. Lett. {\bf 21} 593 (1993).
\bibitem{roz95} Rozenberg M., et al.,
Phys. Rev. Lett. {\bf 75} 105 (1995).
\bibitem{2} Didukh L.,
Ukrainian--French Simposium "Condensed Matter: Science
and Industry". Abstracts, p.275 (Lviv, 1993).
\bibitem{22} Didukh L., Preprint of Institute for Condensed Matter
Physics, ICMP-96-20U, p.~32  (Lviv, 1996, in Ukrainian).
\bibitem{did97jps} Didukh L.,
Journ. of Phys. Stud. {\bf 1} 241 (1997, in Ukrainian).
\bibitem{5} Aligia A.~A., Arrachea L., and Gagliano E.~R., Phys.
Rev.~B. {\bf 51} 13774 (1995).
\bibitem{6} Lin H.~Q. and Hirsch J.~E., Phys.
Rev.~B. {\bf 52} 16155 (1995).
\bibitem{7} Gagliano E.~R., et al.,
Phys. Rev.~B. {\bf 51} 14012 (1995).
\bibitem{stas} Grygorchuk R.~A., Stasyuk I.~V.,
Ukr. Phys. Journ. {\bf 25} 404 (1980, in Russian).
\bibitem{apr} Didukh~L., Cond. Matter Phys.
{\bf 1} 125 (1998).
\bibitem{bari} Bari~R.~H., Adler~D., Lange~ R.~V., Phys. Rev. B.
{\bf 2} 2898 (1970).
\bibitem{pre} Didukh~L. et al, Preprint of Institute for Condensed Matter
Physics  ICMP-03-31U, p.30 (Lviv, 2003, in Ukrainian).
\bibitem{spal90} Spa\l ek~J., Gopalan~P.,
Phys. Rev. Lett. {\bf 64} 2823 (1990).
\bibitem{uhri95} Uhrig~G., Vollhardt~D., Phys. Rev. B.
{\bf 52} 5617 (1995).

\bibitem{ok05} L.~Didukh and O.~Kramar, to be published in Condens. Matter
Phys. (2005).
\bibitem{flou05} A.McCollam et al, Physica B  {\bf 359-361},
p.1-8 (2005).

\end{thebibliography}
\end{document}